\documentclass[epj]{svjour} 
 
\usepackage{graphicx} 
\usepackage{fancyhdr} 
\usepackage{color} 
\usepackage{epsfig} 
\usepackage{amsfonts,amssymb} 
\def\makeheadbox{{
 }} 
\def\mytitle{My title}  
\def\myauthors{My name}   
\def\mytype{My type of session} 
\def\mysession{My session} 

\def\mytitle{Detection of long-lived $\stau s$ and $\sG s$ at the ILC}  
\def\myauthors{H-U Martyn}    
\def\mytype{Contributed Talk}     
\def\mysession{Colliders - SUSY Phenomenology} 
 
\newcommand{\eV}  {\mathrm{eV}} 
 
\newcommand{\MeV} {\mathrm{MeV}} 
\newcommand{\GeV} {\mathrm{GeV}} 
\newcommand{\TeV} {\mathrm{TeV}} 
\newcommand{\fb}  {\mathrm{fb}} 
\newcommand{\fbi} {\mathrm{fb}^{-1}} 
\newcommand{\cm}  {\mathrm{cm}}
\newcommand{\sek} {\mathrm{s}} 
\newcommand{\msek}{\mathrm{ms}} 
\newcommand{\nsek}{\mathrm{ns}} 
\newcommand{\cO}  {{\cal O}} 
\newcommand{\cL } {{\cal L}} 
 
\newcommand{\cB } {{\cal B}} 
\def\susy   {{\sc Susy}} 
\def\ilc    {{\sc ILC}} 
 
\def\ee    {e^+e^-} 
\def\ti    {\tilde} 
\def\sa    {{\ti a}} 
\def\stau  {{\ti\tau}} 
\def\sell  {{\ti\ell}} 
\def\cx    {{\ti\chi}} 
 
\def\nt    {{\ti\chi^0}} 
 
\def\sG    {{\ti G}}

\def\smur  {{\ti\mu}_R} 
 
\def\sel   {{\ti e}_L} 
\def\ser   {{\ti e}_R} 
 
\newcommand{\nn}{\nonumber} 
\newcommand{\eq}[1]{(\ref{#1})} 
\newcommand{\fig}[1]{Fig.~\ref{#1}} 
\newcommand{\tab}[1]{Table~\ref{#1}}

\begin{document} 
\title{Detection of long-lived staus and gravitinos at the ILC} 

\author{Hans-Ulrich Martyn\inst{1,2}                                 
}                     
\institute{I. Physikalisches Institut, RWTH Aachen, Aachen, Germany 
  \and Deutsches Elektronen-Synchrotron DESY, Hamburg, Germany 
} 
%
\date{} 
\abstract{   
  A study is presented illustrating the excellent potential of future  
  International Linear Collider (\ilc) experiments 
  to detect metastable staus $\stau$, measure precisely their mass and
  lifetime, and to determine the mass of the gravitino $\sG$ from the decay
  $\stau\to\tau\sG$, thus providing direct access to the gravitational
  coupling, respectively Planck scale. 
%
} 
\maketitle 

\section{Introduction} 
  \label{intro}  
 
Supersymmetry (\susy) provides an attractive scenario to account for the 
amount of dark matter in the universe. 
If $R$-parity is conserved, the lightest supersymmetric particle (LSP) is 
stable and an ideal dark matter candidate. 
A very interesting option is the spin $3/2$ gravitino $\sG$.   
The mass of the gravitino is set by the \susy\ breaking scale $F$ via      
$  m_{3/2} = m_{\sG}  =  {F}/{\sqrt{3}\, M_P}$, 
with $M_P \simeq 2.4 \cdot 10^{18}~\GeV$ the reduced Planck scale. 
In general $m_{3/2}$ is a free parameter and may 
extend over a wide range of $\cO(\eV-\TeV)$
for gauge, gaugino and supergravity mediated symmetry breaking. 
 
A gravitino LSP may be produced in decays of \susy\ particles. 
If the next-to-lightest supersymmetric particle (NLSP) is 
the scalar tau $\stau$, the dominant process is $\stau\to\tau\sG$. 
Since the coupling is gravitational, the lifetime may be very long,  
ranging from seconds to years. 
The decay-width $\Gamma_\stau$, respectively lifetime 
$\tau =\Gamma_\stau^{-1}$, of the $\stau$ NLSP 
\begin{eqnarray} 
  \Gamma_{\stau\to \tau\sG}  & = & \frac{1}{48\pi M_P^2} 
  \frac{m_{\stau}^5}{m_{\sG}^2} 
  \left[1 -\frac{m_{\tilde{G}}^2}{m_{\stau}^2} \right]^4 \  
  \label{eq:decaywidth} 
\end{eqnarray} 
depends only on the masses $m_\stau$ and $m_\sG$ as well as 
on the Planck scale $M_P$ -- no further \susy\ parameters are required. 
 
The cosmological production of gravitino dark mat\-ter proceeds  
essentially via thermal production and/or late decays of the NLSP. 
The big bang nucleosynthesis puts 
constraints on the $\stau$ lifetime~\cite{steffen}, {\it e.g.} 
from the energy release in hadronic decays one expects
$\tau \lesssim 10^7\,\sek$ for $m_\sG \sim 100\,\GeV$.
Bound states of ${\cal N}\stau^-$ may alter the production of light elements  
considerably, although possible consequences are controversial. 
To avoid $^6$Li overproduction the $\stau$ lifetime should be restricted 
to $\tau\lesssim 5 \cdot 10^3\,\sek$~\cite{pospelov}, 
while the author of~\cite{jedamzik} argues that the  
prior synthesised elements may be destroyed again at longer lifetimes. 
 
Experiments at the \ilc\  
offer a unique possibility to detect long-lived staus 
and to study the properties of gravitinos, 
which otherwise cannot be observed in astrophysical experiments. 
A variety of spectra and \susy\ breaking scenarios have been investigated
experimentally in detail~\cite{martyn};
here just two models, mSUGRA and GMSB scenarios, are presented.

\section{{\boldmath $\stau$} detection \& measurement principles} 
\label{detection} 
 
\begin{figure}[htb] \centering 
  \includegraphics[width=.45\textwidth,angle=0] {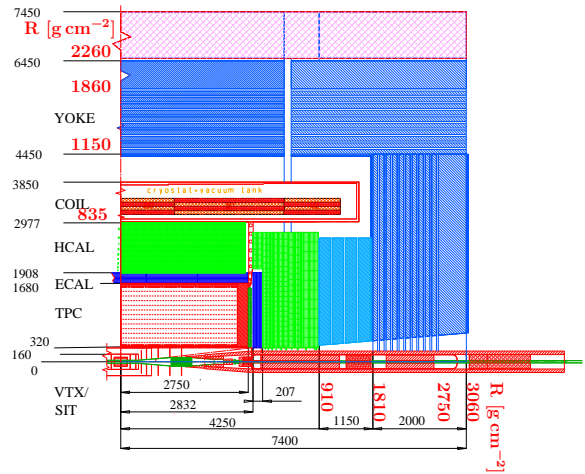} 
  \caption{Quadrant of a typical \ilc\ detector~\cite{tdr}, 
    length units in mm; 
    amount of material indicated by $R~[{\rm g}\,\cm^{-2}]$} 
  \label{fig:detector} 
\end{figure}

A typical \ilc\ detector~\cite{tdr} is displayed in \fig{fig:detector}.
The main characteristics, relevant to the present study, are: 
a TPC with excellent tracking and dE/dx resolution 
to identify slow, heavy particles by ionisation; 
a highly segmented hadronic calorimeter (HCAL) with energy resolutions 
$\delta E_{h}/E = 0.5/\sqrt{E / \GeV}$ for hadrons and 
$\delta E_{em}/E = 0.2/ \sqrt{E / \GeV}$ for electrons/photons; 
an instrumented iron yoke to allow for muon detection and coarse 
calorimetric measurements of hadrons.    
The amount of material available to absorb a heavy $\stau$
in the HCAL or yoke corresponds to an acceptance for scaled momenta of  
$p/m = \beta\gamma\lesssim 0.4-0.5$. 

The stau detection and measurement principle consists of several steps: 
identify a $\stau$ and determine its mass from kinematics;  
follow the track until it is trap\-ped inside the detector; 
observe the stopping point until a decay $\stau\to\tau\sG$ 
is triggered by a large energy release uncorrelated to beam collisions; 
record the decay time to determine the $\stau$ lifetime; 
finally, measure the $\tau$ recoil energy to get the 
gravitino mass 
\begin{eqnarray} 
   E_\tau & = & \frac{m_\stau}{2}\,  
   \left ( 1 - \frac{m_\sG^2 - m_\tau^2}{m_\stau^2} \right ) \ . 
   \label{eq:etau} 
\end{eqnarray} 
 
The \ilc\ provides a very favourable environment. 
The  centre of mass energy can be adjusted to optimise the number of 
observable staus. 
The $\ee$ beams collide in bunch trains of 1~ms duration repeated every 
200~ms; thus,
the detector is most of the time inactive and in principle ideally suited 
to measure long-lived particles.  
However, it is envisaged to operate the HCAL in a pulsed  
mode, switching on only during collisions.
Clearly this concept has to be revised to reach a reasonable duty cycle.

\section{Experimental analyses -- case studies} 
\label{analyses} 
 
The analysis is based on a complete event simulation including QED radiation,
beamstrahlung and detector resolutions. The experimental signature is very
clean and distinct from Standard Model background which can
be efficiently rejected. 
There are no missing particles (except $\nu's$ from decays), the observed
particle momenta are balanced, $| \sum_i \vec{p_i}  | \simeq 0$,
but their moduli don't sum up to the cms energy $\sum_i p_i < \sqrt{s}$.
These features allow the sparticle masses and decay chains to be reconstructed
from the event kinematics.
Each \susy\ event contains two $\stau's$, easily identified by ionisation in
the TPC ($dE/dx\sim\beta^{-2}$), and their passage through the 
detector can be accurately followed. 
The location of stopping $\stau's$ can be determineded within 
a volume of a few $\cm^3$.

The production of low momentum $\stau's$ with a suitable $\beta\gamma$ factor 
to be trapped in the detector proceeds either directly or via cascade decays 
from light sleptons or neutralinos. 
All these processes  
--- $\stau_1\stau_1$, $\ser\ser$, $\smur\smur$ and $\nt_1\nt_1$ --- 
rise only slowly above kinematic threshold with cross sections  
$\sigma\propto\beta^3$, thus providing relatively low rates. 
More efficient, if kinematically accessible, is  associated 
selectron production $\ee\to\ser\sel$, 
increasing as $\sigma\propto\beta$ near threshold.
The event signatures are multi-lepton topologies:
$2\stau_1$ from pair production, $2\stau_1 2\tau$ from neutralino production
and $2\stau_1 2\tau 2\ell$ from selectron and smuon production.

\subsection{mSUGRA scenario {\boldmath GDM~$\epsilon$} } 
\label{gdme} 

In {\it supergravity mediated symmetry breaking} (SUGRA) the gravitino mass
$m_{3/2}$ is a free parameter of the same order as the other sparticle
masses. 
In minimal versions with the $\stau$ NLSP 
the common scalar mass $m_0$ 
has to be small and much lower than the common gaugino mass $M_{1/2}$.
In the mSUGRA scenario GDM ~$\epsilon$~\cite{gdm} a tighter definition is used
with unified scalar and gravitino masses 
$m_0=m_{3/2}=20\,\GeV$, $M_{1/2}=440\,\GeV$,
$A_0=25\,\GeV$, $\tan\beta=15$ and sign\,$\mu = +$.
The corresponding sparticle spectrum is compiled in \tab{tab:gdmspectra}.

\begin{table} \centering  
  \caption{Sparticle masses and decay modes  
    of mSUGRA scenario GDM $\epsilon$ accessible at $\sqrt{s}=500\,\GeV$} 
  \label{tab:gdmspectra} 
  \begin{tabular}{l c l l c l} 
    \noalign{\smallskip}\hline\noalign{\smallskip} 
                & m~[GeV] & $\cB$ ~~~~~~~~&    
                & m~[GeV] & $\cB$  
    \\ \noalign{\smallskip}\hline\noalign{\smallskip} 
    $\stau_1$   & 157.6 & $\tau\sG$    & 
    $\smur  $   & 175.1 & $\mu\tau\stau$ \\ 
    $\ser  $    & 175.1 & $e\tau\stau$ & 
    $\sel  $    & 303.0 & $e\nt_1$     \\ 
    $\nt_1 $    & 179.4 & $\tau\stau$  & 
    $\sG  $     &  20   &     \\ 
    \noalign{\smallskip}\hline 
  \end{tabular}     
\end{table} 
  
The experimental assumptions for the case study are 
the canonical \ilc\ energy $\sqrt{s}=500\,\GeV$ and an integrated luminosity 
$\cL= 100\,\fbi$ (less than 1 year of data taking).
The inclusive $\stau$ production cross section is
$\sigma(\stau_1\stau_1 X) = 300 \,\fb$.

\begin{figure*} \centering 
  {\bf\boldmath mSUGRA scenario GDM $\epsilon$ } \\[1.8ex] 
    \includegraphics[width=1.\textwidth]{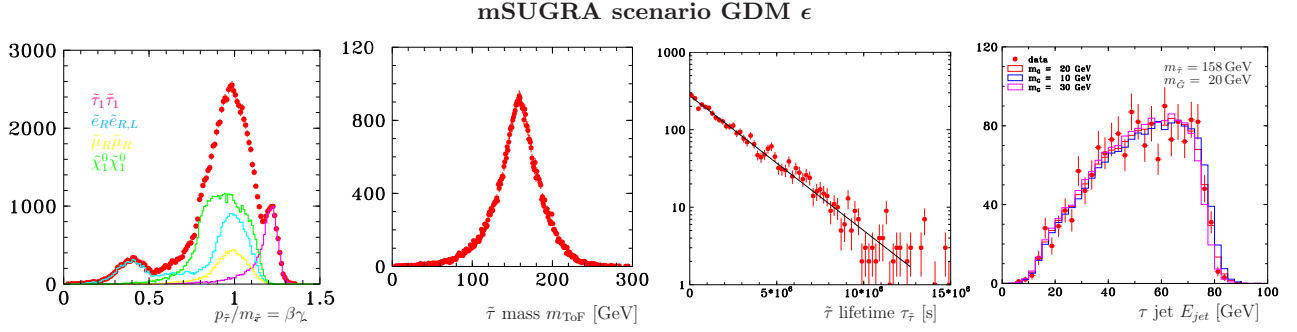} 
  \caption{ GDM $\epsilon$ scenario,  
    assuming $\cL = 100\,\fbi$ at $\sqrt{s} = 500\,\GeV$: 
    (a) 
    $\stau$ production spectra of scaled momentum $p/m = \beta\gamma$ 
    with contributions from various processes;
    (b)
    $\stau$ mass $m_{\rm ToF}$ spectrum; 
    (c) 
    $\stau$ lifetime distribution;
    (d)
    $\tau$ jet energy spectrum of the decay $\stau_1\to\tau\sG$,
    compared with simulations of $m_\sG = 20\,\GeV$, 
    $10\,\GeV$ and $30\,\GeV$ } 
  \label{fig:gdmepsilon}       
\end{figure*}

The prolific {\it stau production} rate is characterised by
the scaled momentum distribution $p/m=\beta\gamma$, shown in 
\fig{fig:gdmepsilon}\,a for the various reactions.
The majority of particles come from diagonal slepton and neutralino pairs and 
leave the detector (peak around $\beta\gamma\simeq 1$).  
One observes, however, a second peak at low $\beta\gamma\lesssim 0.5$ 
from cascade decays of $\ser\sel$ production, which will be stopped in the
detector. 
The number of $\stau's$ trapped are $N_\stau^{\rm hcal} = 4100$ 
and $N_\stau^{\rm yoke} = 1850$ in the hadron calorimeter 
and yoke, respectively. 
The choice of energy turns out to be optimal.
Similar rates can be obtained at $380\,\GeV$, just above
the slepton/neutralino thresholds.
Selecting $\stau_1\stau_1$ pairs to absorb $\stau's$ is much less efficient,
{\it e.g.} $N_\stau^{\rm hcal} = 1600$ at $\sqrt{s}=340\,\GeV$ 
for the same integrated luminosity.
 
The {\it stau mass} measurement is based on the kinematics of 
$e^+e^-\to\stau_1\stau_1$, see magenta curve in \fig{fig:gdmepsilon}~a,  
to be identified as a pair of collinear, non-interacting particles with  
momenta  $p_\stau<\sqrt{s}/2 = E_\stau$. 
A determination of the mean momentum 
$\langle p_\stau \rangle = 192.4 \pm 0.2\;\GeV$  
leads to a precise $\stau$ mass of  
\begin{eqnarray} 
  m_\stau &=& 157.6\pm 0.2\;\GeV\ . \nn 
\end{eqnarray} 
 
Alternatively one may use the much larger sample of all identifed $\stau's$,
also those leaving the detector, and perform a the time-of-flight 
measurement using the calorimeter.
The time resolution is
$\delta t = 1\,\nsek$, the track lengths are $\sim 2- 4$\,m.
The reconstructed mass distribution 
$m_{\mathrm ToF} = \sqrt{(1/\beta^2-1)\,p^2}$,
displayed in \fig{fig:gdmepsilon}~b,
provides an accuracy $\delta m_{\mathrm ToF} = 0.15\;\GeV$, 
similar to that of the momentum measurement.

The {\it stau lifetime} measurement is based on the decays of $\stau's$ 
which have been stopped in the detector.  
Requiring an isolated energetic cluster or muon above a certain threshold 
originating somewhere inside the sensitive 
fiducial volume of the calorimeter or yoke, 
results in the decay time distribution shown in \fig{fig:gdmepsilon}~c. 
A fit to the spectrum gives a $\stau$ lifetime of  
\begin{eqnarray} 
  \tau &=& (2.6\pm0.05)\cdot10^6\,\sek \ , \nn 
\end{eqnarray} 
corresponding to roughly one month. 
For the actual conditions the lifetime scales as $\tau \sim m^2_\sG$
(also valid for very low masses)
and rises significantly faster for $m_\sG \gtrsim 0.25\,m_\stau$,  
reaching $10^8\,\sek$ for a $75\,\GeV$ gravitino. 

{\it Note:}  
The relative precision on the $\stau$ lifetime does  
not depend on the gravitino mass, should it be much lighter 
as for larger mass splittings or in gauge mediated supersymmetry models.  
Technically, there may be a limitation to measure lifetimes below $1\,\msek$ 
which corresponds to the sensitive time of beam collisions.

A {\it direct gravitino mass} measurement can be performed by exploiting  
the $\tau$ recoil of the decay $\stau\to\tau\sG$, see \eq{eq:etau}.
The upper endpoints of the energy spectra  
which coincide with the primary $\tau$ energy $E_\tau=77.5\,\GeV$, 
are directly related to the masses involved. 
The leptonic 3-body decays $\tau\to \ell\nu\nu$ are not very useful due to 
the soft spectrum peaking at low values. 
Well defined upper edges are provided by the hadronic decays to heavier final  
states $\tau\to\rho\nu$ and $\tau\to\pi\pi\pi\nu$. 
The energy distribution of both decay modes, defined as `$\tau$ jets', 
is shown in \fig{fig:gdmepsilon}~d.  
In order to illustrate the sensitivity to the gravitino mass,
simulations assuming the nominal value of $m_\sG=20\,\GeV$ and shifted 
by $\pm 10\,\GeV$ are shown as well.  
A fit to the $\tau$ jet energy spectrum, yields a gravitino mass 
\begin{eqnarray}   
  m_\sG&=&20\pm 4\,\GeV \ . \nn 
\end{eqnarray} 

Combining all results one can test the gravitational coupling of  
the stau to the gravitino and access the Planck scale,  
respectively Newton's constant. 
Inserting the expected values and accuracies 
on $m_\stau$, $\tau$ and $m_\sG$ in \eq{eq:decaywidth} 
one finds for the supergravity Planck scale 
\begin{eqnarray}   
  M_{P} & = & (2.4 \pm 0.5)\cdot 10^{18}\;\GeV \ , \nn 
\end{eqnarray} 
where the error is dominated by the precision on the gravitino mass.
 
The {\it gravitino mass} can be deduced more precisely 
from the $\stau$ mass and lifetime, 
if the gravitational coupling is shown to be responsible for the decay
or is {\it assumed} and the macroscopic value of 
$M_P$ is taken in the decay-width of \eq{eq:decaywidth}. 
The resulting gravitino mass is $m_\sG=20\pm 0.2\,\GeV$,  
where the error is dominated by the lifetime measurement. 
This mass value can be used to get access to
the supersymmetry breaking scale $F=\sqrt{3}\,M_P\,m_{3/2} 
 = (8.3\pm0.1)\cdot10^{19}\,\GeV^2$,
which is an important parameter to unravel the nature of the supersymmetry 
breaking mechanism.

The expected accuracies on the $\stau$ and $\sG$ observables 
and derived quantities of the GDM~$\epsilon$ scenario are summarised in
\tab{tab:results}. 
 
\renewcommand{\arraystretch}{1.1} 
\begin{table} [htb] 
  \centering 
  \caption{Expected accuracies on  $\stau$ and $\sG$ 
    properties of observables and derived quantities;  
    GDM $\epsilon$ scenario, based on  
    $\cL=100\,\fbi$ at $\sqrt{s}=500\,\GeV$} 
  \label{tab:results} 
  \begin{tabular}{l l l l l l } 
    \multicolumn{2}{l}{observables} & 
    \multicolumn{2}{l}{derived quantities} 
    \\ \noalign{\smallskip}\hline\noalign{\smallskip} 
    $m_\stau$    & $ 157.6 \pm 0.2\,\GeV$      \\ 
    $\tau_\stau$ & $ (2.6\pm0.05)\,10^6\,\sek$  & 
    $m_{\sG}^{t_{\stau}}$ & $ 20 \pm 0.2 \,\GeV$ \\ 
    $m_{\sG}^{E_\tau}$ & $ 20 \pm 4.0 \,\GeV $ 
                       & $M_P$ & $ (2.4\pm0.5)\, 10^{18}\,\GeV$ \\ 
  \end{tabular}   
\end{table}

It is a unique feature of gravitino LSP scenarios that the Planck scale can  
be directly measured in microscopic particle experiments by studying 
the properties of the NLSP and its decay. 
A further interesting test to reveal the nature of the gravitino 
as the supersymmetric partner of the graviton 
would be to determine the spin,  
which is possible by studying correlations in the radiative decays 
$\stau\to\tau\gamma\sG$~\cite{buchmueller}.

\begin{figure*} 
  \centering 
  {\bf \boldmath  GMSB scenario SPS 7 } \\[1.8ex] 
  \includegraphics[width=0.75\textwidth,angle=0]{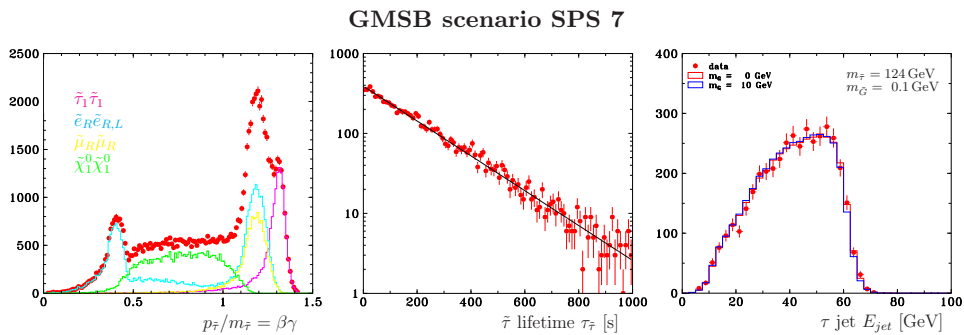}  \\
  \caption{ SPS 7 scenario,  
    assuming $\cL = 100\,\fbi$ at $\sqrt{s} = 410\,\GeV$: 
    (a) 
    $\stau$ production spectra of scaled momentum $p/m = \beta\gamma$ 
    with contributions from various processes;
    (b) 
    $\stau$ lifetime distribution;
    (c)
    $\tau$ jet energy spectrum of the decay $\stau_1\to\tau\sG$
    compared with simulations of $m_\sG = 0\,\GeV$ and $10\,\GeV$ } 
  \label{fig:sps7}       
\end{figure*}

\subsection{GMSB scenario SPS 7} 
  \label{gmsb_sps7}       

{\it Gauge mediated symmetry breaking} (GMSB) 
usually occurs at rather low scales and a light gravitino is naturally 
the LSP. 
Typical masses are of order eV to keV which may be extended in the GeV range. 
The GMSB reference scenario SPS~7~\cite{sps} 
is described by the conventional parameters
$\Lambda=40\,\TeV$, $M_m=80\,\TeV$, $N_m=3$, $\tan\beta=15$ and
sign\,$\mu = +$,
The  sparticles are relatively light:
    $m_{\stau_1}   = 123.4 \,\GeV$,
    $m_{\sell_R}   = 130.9 \,\GeV$,
    $m_{\sell_L}   = 262.8 \,\GeV$,
    $m_{\nt_1}     = 163.7 \,\GeV$.
The gravitino mass is set arbitrarily to $m_{\sG}=0.1\,\GeV$.
 
The SPS~7 model is investigated assuming $\sqrt{s}=410\,\GeV$ 
and $\cL=100\,\fbi$, with a large inclusive $\stau$ cross section of 
$\sigma(\stau_1\stau_1 X) = 420 \,\fb$.
As seen in the $\beta\gamma$ distribution of \fig{fig:sps7}~a, 
most $\stau's$ leave the detector. 
There is, however, a large signal at $\beta\gamma\simeq 0.4$  
from $\ser\sel$ production, contributing to samples of  
 $N_\stau^{\rm hcal} = 10000$ and
 $N_\stau^{\rm yoke} = 4900$
trapped $\stau's$ in the calorimeter and yoke, respectively. 

The analysis of $\stau_1\stau_1$ pair production yields  
a mass of $m_{\stau_1}=124.3\pm 0.1\,\GeV$.
From a fit to the decay time distribution, shown in \fig{fig:sps7}~b,
one obtains a lifetime of $\tau=209.3\pm \,2.4\,\sek$. 
These values can be used to derive a very accurate gravitino mass of  
$m_\sG=100\pm 1\,\MeV$ {\it assuming} a gravitational coupling.
To illustrate the sensitivity to low gravitino masses
as expected in many GMSB models:
a gravitino mass of $0.5\,\MeV$ corresponds
a $\stau$ lifetime of $5\,\msek$, which should be easily measurable.

The $\tau$ jet recoil energy spectrum is displayed in \fig{fig:sps7}~c. 
As can be seen from the simulation curves for 0\,GeV and 10\,GeV
gravitinos, the measurement is not sensitive to such low masses
and can only serve to set an upper limit 
of $m_\sG < 9\,\GeV$ (at 95\% CL). 
 
This limitation can be understood from \eq{eq:etau}. 
The sensitivity to low gravitino masses decreases rapidly, {\it e.g.} 
for $m_\sG = 0.1\,m_\stau$ 
the energy deviates from $m_{\stau}/2$ by one percent,
which requires $E_\tau$ to be known with a precision well below.
Notice that $E_\tau$ is the maximum, the upper edge
of the observable $\tau$ jet energy.
A direct measurement of very large $\stau - \sG$ mass splittings 
becomes experimentally extremely difficult, getting gradually impossible
for ratios $m_\sG/m_\stau \lesssim 0.1$.

The nature of the LSP remains undetermined without knowing the gravitino mass.
Further information can be gained from a study of
radiative three-body decays $\stau\to\tau\gamma\sG$.
The differential decay rates and $\gamma - \tau$ correlations
for a light spin $3/2$ gravitino $\sG$ to be compared with 
a spin $1/2$ neutralino $\cx$~\cite{buchmueller}
and a spin $1/2$ axino $\sa$~\cite{brandenburg} have been calculated
and found to be quite different
in particular phase space regions of back-to-back topologies.
Experimentally the analysis of radiative $\stau$ decays
is quite ambitious:
the branching ratios are suppressed by two orders of magnitude
and single $\gamma's$ have to be disentangled from 
the bulk of photons and hadrons in $\tau$ decays. 
The excellent performance of the proposed
high granularity, 'pictorial' calorimeter~\cite{tdr} 
together with the large data samples expected at the \ilc\
should make it possible to discriminate between a light gravitino, 
a neutralino and an axino LSP.

\section{Conclusions}

Future \ilc\ experiments have a rich potential to 
study \susy\ scenarios
where the gravitino $\sG$ is the LSP
and a charged stau $\stau$ is the long-lived, metastable NLSP.
Precise determinations of the $\stau$ mass and lifetime 
and of the $\sG$ mass appear feasible 
already with moderate integrated luminosity. 
(More \susy\ scenarios can be found in \cite{martyn}.)
A  measurement of the gravitino mass 
from the $\tau$ recoil spectra of the  decay $\stau\to\tau\sG$ 
gives access to the gravitational coupling, 
{\it i.e.} to the Planck scale, 
and thus provides a unique test of supergravity.
Such observations will put stringent constraints on 
an interpretation of the gravitino as dark matter candidate.


\begin{thebibliography}{999} 
   
\bibitem{steffen} 
  F.~D.~Steffen, 
  AIP Conf.\ Proc.\  {\bf 903} (2007) 595. 
 
\bibitem{pospelov} 
  M.~Pospelov, 
  Phys.\ Rev.\ Lett.\  {\bf 98} (2007) 231301. 
 
\bibitem{jedamzik} 
  K.~Jedamzik, 
  arXiv:0707.2070 [astro-ph]. 

 \bibitem{martyn} 
  H.-U.~Martyn, 
  Eur.\ Phys.\ J.\  C {\bf 48} (2006) 15. 
 
\bibitem{tdr} {\sc Tesla} Technical Design Report, DESY 2001-011, 
        Part IV: {\em A Detector for TESLA}.    

\bibitem{gdm} 
  A.~De Roeck, 
  J.~R.~Ellis, F.~Gianotti, F.~Moortgat, K.~A.~Olive, L.~Pape, 
  Eur.\ Phys.\ J.\  C {\bf 49} (2007) 1041. 
 
\bibitem{buchmueller}
  W.~Buchm\"uller, K.~Hamaguchi, M.~Ratz, T.~Yana\-gida,
  Phys.\ Lett.\ B {\bf 588} (2004) 90.
 
\bibitem{sps} B.C.~Allanach {\it et al.}, 
  Eur.\ Phys.\ J.\ C {\bf 25} (2002) 113. 
        
\bibitem{brandenburg}
  A.~Brandenburg, 
  L.~Covi, K.~Hamaguchi, L.~Rosz\-kowski, F.~D.~Steffen,
  Phys.\ Lett.\ B {\bf 617} (2005) 99.
  
\end{thebibliography}
\end{document}